\documentclass[preprintnumbers, floatfix, preprintnumbers, letterpaper, twocolumn, superscriptaddress,nofootinbib]{revtex4-2}
\pdfoutput=1 
\usepackage{graphicx}
\usepackage{microtype}
\usepackage{amsmath}
\usepackage{amssymb}
\usepackage{subfigure}
\usepackage{url}
\usepackage{hyperref}
\usepackage{bigints}

\usepackage{xcolor}
\usepackage{color}
\usepackage{mathrsfs}
\usepackage{calrsfs}
\usepackage{amsfonts}
\usepackage{tabularx}
\usepackage{latexsym}
\usepackage{ragged2e}
\usepackage{epsfig}
\usepackage{textcomp}
\usepackage{float}
\newtheorem{theorem}{Theorem}

\newtheorem{conjecture}{Conjecture}

\usepackage{caption}
\DeclareCaptionJustification{justified}{\leftskip=0pt \rightskip=0pt \parfillskip=0pt plus 1fil}
\definecolor{vividviolet}{rgb}{0.62, 0.0, 1.0}
\definecolor{amaranth}{rgb}{0.9, 0.17, 0.31}
\definecolor{palatinateblue}{rgb}{0.15, 0.23, 0.89}
\definecolor{brightpink}{rgb}{1.0, 0.0, 0.5}
\definecolor{cornflowerblue}{rgb}{0.39, 0.58, 0.93}
\definecolor{deepcarminepink}{rgb}{0.94, 0.19, 0.22}
\definecolor{radicalred}{rgb}{1.0, 0.21, 0.37}

\hypersetup{ linktoc=all,
	colorlinks, linkcolor={palatinateblue},
	citecolor={brightpink}, urlcolor={amaranth}
}
\newcommand{\changeurlcolor}[1]{\hypersetup{urlcolor=#1}}
\graphicspath{{Images/}}

\usepackage{tipa}
\newcommand{\arc}[1]{{%
  \setbox9=\hbox{#1}%
  \ooalign{\resizebox{\wd9}{\height}{\texttoptiebar{\phantom{A}}}\cr#1}}}
\def\sideremark#1{\ifvmode\leavevmode\fi\vadjust{\vbox to0pt{\vss% the remark
			\hbox to 0pt{\hskip\hsize\hskip1em%                          will appear only
				\vbox{\hsize1.3cm\tiny\raggedright\pretolerance10000%          on the side
					\noindent #1\hfill}\hss}\vbox to8pt{\vfil}\vss}}}%
\def\beq{\begin{equation}}
\def\eeq{\end{equation}}

\begin{document}
\title{Black Holes, Complex Curves, and Graph Theory:\\ Revising a Conjecture by Kasner}

\author{Yen Chin \surname{Ong}}
\email{ycong@yzu.edu.cn}
\affiliation{Center for Gravitation and Cosmology, College of Physical Science and Technology, Yangzhou University, \\180 Siwangting Road, Yangzhou City, Jiangsu Province  225002, China}
\affiliation{Shanghai Frontier Science Center for Gravitational Wave Detection, School of Aeronautics and Astronautics, Shanghai Jiao Tong University, Shanghai 200240, China}

\begin{abstract}
The ratios $\sqrt{8/9}=2\sqrt{2}/3\approx 0.9428$ and $\sqrt{3}/2 \approx 0.866$ appear in various contexts of black hole physics, as values of the charge-to-mass ratio $Q/M$ or the rotation parameter $a/M$ for Reissner-Nordstr\"om and Kerr black holes, respectively. In this work, in the Reissner-Nordstr\"om case, I relate these ratios with the quantization of the horizon area, or equivalently of the entropy. Furthermore, these ratios are related to a century-old work of Kasner, in which he conjectured that certain sequences arising from complex analysis may have a quantum interpretation. These numbers also appear in the case of Kerr black holes, but the explanation is not as straightforward. The Kasner ratio may also be relevant for understanding the random matrix and random graph approaches to black hole physics, such as fast scrambling of quantum information, via a bound related to Ramanujan graph. Intriguingly, some other pure mathematical problems in complex analysis, notably complex interpolation in the unit disk, appear to share some mathematical expressions with the black hole problem and thus also involve the Kasner ratio.
\end{abstract} 

\maketitle
\section{Introduction: $2\sqrt{2}/3$ and $\sqrt{3}/2$}

In my previous work \cite{2309.04110}, motivated by the maximum force conjecture \cite{0210109,1408.1820,0607090,724159} of general relativity, I investigated the quantity $F_\text{therm}:=\partial M/\partial r_+$, which I referred to as the ``thermodynamic force'' of a black hole (it has the correct physical dimension of a force; see also \cite{1809.00442}). Here $r_+$ denotes the event horizon and $M \equiv M(Q,r_+)=(Q^2+r_+^2)/2r_+$ is the mass function of a Reissner-Nordstro\"m black hole\footnote{In the units in which $c=G=\hbar=k_B=4\pi\epsilon_0=1$.}. Similarly, for the Kerr case one has $M(a,r_+)=(a^2+r_+^2)/2r_+$. 

In terms of $x:=Q/M$ (or $x:=a/M$ for Kerr black holes), the thermodynamic force reads
\begin{equation}
F_\text{therm}=\frac{\sqrt{1-x^2}}{1+\sqrt{1-x^2}}.
\end{equation} 
We have assumed that both $Q$ and $a$ are non-negative.
The maximum force conjecture $F \leqslant 1/4$ applied to $F_\text{therm}$ implies that $x \leqslant 2\sqrt{2}/3$. Though the analysis so far is based on the maximum force conjecture, this specific value of the charge-to-mass ratio being somewhat special is not. One can find various occurrences of this number in the literature concerning Reissner-Nordstr\"om black holes, for example:
\begin{itemize}
\item[(1)] The $(1+1)$-dimensional stress-energy tensor expectation $\left\langle T_t^{~t}(r) \right\rangle$ changes sign at $x=2\sqrt{2}/3$ \cite{9607048}.
\item[(2)] A freely falling observer would not detect any Hawking radiation near the black hole horizon once $x \geqslant 2\sqrt{2}/3$ \cite{{0805.1876}}.
\item[(3)] The frequency of a typical Hawking particle near horizon becomes negative for $x \geqslant 2\sqrt{2}/3$ \cite{2003.10429}. (The technique used is based on the fact that, classically, the radial tidal force changes sign for this range of $x$ \cite{1602.07232}.)
\item[(4)] A suitably defined effective temperature at the horizon is negative for $x \geqslant 2\sqrt{2}/3$ \cite{2301.12319}.
\item[(5)] Smooth embedding for the $(t,r)$-part of Reissner-Nordstr\"om geometry into a (2+1)-dimensional Minkowski spacetime fails if $Q/M \geqslant 2\sqrt{2}/{3}$ \cite{0305102}.
\end{itemize}
In other words, a variety of techniques in (1)-(4) indicate that something peculiar happens to the properties of Hawking radiation once the charge-to-mass ratio is sufficiently large (we will return to the Kerr case later in Sec.(\ref{kerr})), so the result is likely robust. Points (3) and (5) also show that  $x=2\sqrt{2}/3$ is not just interesting quantum mechanically, but also classically.

Motivated partly by the result in holography, in which asymptotically locally anti-de Sitter black holes with toral topology suffers brane nucleation instability when the charge-to-mass ratio becomes large enough \cite{1012.4056}, and partly by the maximum force conjecture (whose violation might also indicate instability),
I conjectured in \cite{2309.04110} that Reissner-Nordstr\"om black holes become a highly quantum object once its charge-to-mass ratio lies in the ``quantum dominance bound'' $Q/M \geqslant 2\sqrt{2}/3$ regime.
If so, such black holes can no longer be accurately described by semi-classical general relativity\footnote{In the near-extremal case, there are more rigorous arguments \cite{2210.02473,2303.07358,2307.10423}, though for the bound proposed here, 94\% of extremality does not seem like ``near''. Still, brane nucleation \cite{1012.4056} mentioned above already occurs at $x=0.916$ in AdS$_4$ and 0.958 in AdS$_5$, so it is not unheard of that semi-classical description can fail while the black hole is not yet $\varepsilon$ close to being extremality.}. 
However, the calculation therein is completely classical, since a force is necessarily a classical quantity free of $\hbar$. Of course, as emphasized in \cite{2309.04110} this does not mean that a force cannot have a quantum origin -- a Casimir force is an example. Nevertheless, if the claim is that semi-classical general relativity fails, then it should also mean that quantum gravitational correction has become important. Is there a way to link the quantum dominance bound $Q/M \geqslant 2\sqrt{2}/3$ with any quantum gravitational effect?

In this work, I shall argue that the answer is ``yes'', at least for the case of the Reissner-Nordstr\"om black hole. In particular we can link this to the quantization of the horizon area. But before that, in Sec.(\ref{2}) let us take a detour to introduce a seemingly unrelated topic concerning what I call ``Kasner sequence'' -- a little known result in complex analysis\footnote{As per August 2024, according to Google Scholar the first relevant paper of Kasner \cite{K1}, had only been cited 4 times.}. In Sec.(\ref{3}) I will relate area quantization with the Kasner sequence (and Kasner ratio). Interestingly, this not only relates to the occurrence of $2\sqrt{2}/3$, but also that of $\sqrt{3}/2$, which corresponds to the Davies point \cite{davies} of Reissner-Nordstr\"om black holes\footnote{In addition, though an approximate result (since it depends on 2nd order near-horizon expansion), Ref.\cite{2111.06089} indicated that for $Q/M > \sqrt{3}/2$ the so-called ``chaos bound'' involving the Lyapunov exponent near the horizon is locally violated by the null and time-like circular motion with large angular momentum.}.
Then in Sec.(\ref{kerr}) the rotating case of Kerr black holes will be discussed as the scenario there is somewhat different -- the link to area quantization seems to fail, but Kasner sequence still appears in an interesting way. This is followed by a discussion on the ratio of the horizon radii in Sec.(\ref{cc}) and how it relates to the Kasner ratio; I also give a brief comment on cosmic censorship in this section. Next, possible relations to graph theory are discussed in Sec.(\ref{graph}), and some speculations concerning random matrices and quantum information scrambling are made.  Finally, we conclude with more discussions in Sec.(\ref{discuss}), in which we shall also briefly mention the occurrence of $2\sqrt{2}/3$ in the purely mathematical problem of complex interpolation. 

\section{Complex Curves and Kasner Sequence}\label{2}

More than a century ago, mathematician Edward Kasner (who is well-known in the gravity community for ``Kasner metric'' and ``Kasner singularities'') investigated a purely geometrical problem \cite{K1,K2,K3}: consider two points $P$ and $Q$ on a curve, and compute the arclength $\arc{PQ}$ as well as the length of the chord $|PQ|$, what is the ratio of the arc to the chord $\arc{PQ}/|PQ|$ as $P \rightarrow Q$? For real smooth curves the answer is 1. 
If one gives up smoothness and makes the curve sufficiently ``crinkly'' (in Kasner's own word) then the limit may easily exceed 1.
However, Kasner found that the ratio can be \emph{less} than unity if we allow \emph{complex curves}. 

To see this, in the usual Cartesian $(x,y)$-coordinates, we can write down the equation of a curve through $P(0,0)$ in the form of a power series in $x$:
\begin{equation}
y = c_1 x + c_2 x^2 + c_3 x^4 + \cdots,
\end{equation}
where the coefficients $c_i \in \Bbb{C}$. The chord length from $P(0,0)$ to $Q(x,y)$ is simply given by the norm:
\begin{flalign}
|PQ|&=\sqrt{x^2+y^2}\\
&=x\left[1+c_1^2+ 2c_1c_2 x+ (2c_1c_3+c_2^2)x^2 + \cdots\right]^{\frac{1}{2}}.
\end{flalign}
On the other hand, the arclength between the two points is given by the integral
\begin{flalign}
\arc{PQ} &= \bigintsss_0^x \sqrt{1+\left(\frac{dy}{dx}\right)^2} dx \\
&= \int_0^x \left[1+c_1^2 + 4c_1c_2 x + (6c_1c_3 + 4c_2^2)x^2 + \cdots \right]^{\frac{1}{2}} dx
\end{flalign}
In the limit of small $x$, if $c_1 \neq \pm i$, then both $|PQ|$ and $\arc{PQ}$ equal $\sqrt{1+c_1^2}x$, so that
\begin{equation}
\lim_{x \to 0} \frac{\arc{PQ}}{|PQ|} = 1.
\end{equation}
If $c_1 = \pm i$, but $c_2 \neq 0$, then 
\begin{flalign}
\lim_{x \to 0} \frac{\arc{PQ}}{|PQ|} &= \frac{\displaystyle \int_0^x \sqrt{4c_1c_2 x}  + \cdots dx }{\sqrt{2c_1c_2}x^{3/2} + \cdots} \\
&=\frac{\frac{2}{3}\sqrt{4c_1c_2}x^{3/2}+ \cdots}{\sqrt{2c_1c_2}x^{3/2} + \cdots} \\
&=\frac{\frac{2}{3}\cdot 2}{\sqrt{2}}=\frac{2\sqrt{2}}{3}.
\end{flalign}
Kasner provided an explicit example for which the limit of this ratio at the origin is ${2\sqrt{2}}/{3}$ : the complex parabola $y = x^2 + ix$. This can be verified straightforwardly. 

Now we can continue with this process by considering that in addition to $1+c_1^2=0$, also that $c_2 =0$, in which case the limit of the ratio is $\sqrt{3}/2$ and so on.
In this way, if $c_1 = \pm i$, $c_2 = c_3 = \cdots = c_{k-1} = 0$ but $c_k \neq 0$, we would obtain
\begin{flalign}\label{rac}
R_k:=\lim_{x \to 0} \frac{\arc{PQ}}{|PQ|} &= \frac{\frac{2}{k+1}\sqrt{2kc_1c_k}x^{\frac{k+1}{2}}}{\sqrt{2c_1c_k}x^{\frac{k+1}{2}}} \notag \\
&= \frac{2\sqrt{k}}{k+1}.
\end{flalign}
Kasner called this the ``rac curvature'', but I would refer to it as the \emph{Kasner ratio}\footnote{For simplicity, in later discussions, we keep the name ``Kasner ratio'' even when $k$ is not an integer. Of course a sequence only makes sense with $k \in \Bbb{N}$.}. The ``rac set'', or what I will refer to as the \emph{Kasner sequence}, is obtained by running through $k \in \Bbb{N}$, i.e.,
\begin{equation}
\left\{1, \frac{2\sqrt{2}}{3}, \frac{\sqrt{3}}{2}, \frac{4}{5}, \frac{\sqrt{5}}{3}, \frac{2\sqrt{6}}{7}, \frac{\sqrt{7}}{4}, \frac{4\sqrt{2}}{9},\frac{3}{5}, \cdots \right\}, 
\end{equation}
or approximately
\begin{equation}\notag
\left\{1, 0.9428, 0.866, 0.8, 0.7454, 0.6999, 0.6614, 0.6285,0.6, \cdots \right\}.
\end{equation}

We note that the second and third terms are exactly the charge-to-mass ratios of interest as discussed in the Introduction. Interpreted as charge-to-mass ratio the first term of course corresponds to the extremal black hole. Kasner already noted in his \emph{Nature} paper in 1921 \cite{K2} that the Kasner ratio has a ``physical interpretation'', though in special relativity. Namely, if a hypothetical particle starts out with the velocity of light and then continuously slows down (the initial acceleration being nonzero), then the limit of the ratio of the arc-to-chord on the worldline is $R_2=2\sqrt{2}/3$. 

Even more curiously, since $k$ is a positive integer,
Kasner made a ``quantization conjecture'' \cite{K3} even though the whole results obtained thus far are in the context of complex geometry:
\begin{quote}
``The \emph{discontinuity} which thus presents itself in the rac set of rectilinear motion suggests at least an analogy with the discontinuous character
of energy radiation (quantum theory) [$\cdots$] If $R$ denotes the rac in [Eq.(\ref{rac})] we may write
\begin{equation}\label{ls}
\frac{R^2}{4}=\frac{m-1}{m^2}=\frac{1}{m}-\frac{1}{m^2},
\end{equation}
where $m=n+1$\footnote{In Kasner's paper, it was written $m=n-1$; this is a typo.} is an integer greater than one. Perhaps this formula
may in the future be connected with spectrum line series.''
\end{quote}
The emphasis on ``discontinuity'' is his original, which meant the set is discrete. This conjecture is probably based on the appearance of Eq.(\ref{ls}) that is reminiscent of the Balmer series or Rydberg formula in general.

In the following, I will argue that Kasner's foresight is remarkable -- the Kasner ratio does find a correspondence in quantum physics, but instead of spectrum line series of ordinary materials, it is related to quantum gravity, or more precisely, the spectrum of the quantized area (or equivalently, entropy) of Reissner-Nordst\"om black holes.

\section{Relating Reissner-Nordstr\"om Area Quantization to Kasner Ratio}\label{3}

Consider the quantized entropy for Reissner-Nordstr\"om black holes \cite{0102061,0207169,0209039}:
\begin{equation}
S = \pi\left[2n + (p+1)\right], ~~ Q=\sqrt{p},~~ n, p \in \Bbb{N}.
\end{equation}
To avoid other quantum gravitational effects that arise from strong curvature, we only consider large black holes, so that its entropy is given by the area law $S=\pi r_+^2$, with subleading corrections suppressed.
Then we can identify
\begin{equation}
\left(M+\sqrt{M^2-p}\right)^2 = 2n + p + 1,
\end{equation}
which allows us to solve for the mass in terms of the integers $n$ and $p$:
\begin{equation}
M = \frac{1}{2}\frac{2n+p+1}{\sqrt{2n+p+1}}.
\end{equation} 
Consequently, the charge-to-mass ratio is (c.f., however, \cite{1512.03095})
\begin{equation}\label{Q-M}
\frac{Q}{M}=\frac{2\sqrt{p}\sqrt{2n+p+1}}{2n+p+1}.
\end{equation}
We claim that this can be identified with the Kasner sequence. Indeed, if this is the case, then for $m \in \Bbb{N}$,
\begin{equation}
\frac{2\sqrt{m}}{m+1}=\frac{2\sqrt{p}\sqrt{2n+p+1}}{2n+p+1}.
\end{equation}
This implies
\begin{equation}\label{mnumber}
m = \frac{2n+1+p}{p}.
\end{equation}
Surprisingly, this particular expression on the right hand side has appeared before in the literature; it is also equal to the square root of the ratio between the magnitude of the surface gravity of the inner and outer horizons for quantized Reissner-Nordstr\"om black holes \cite{1111.4164}:
\begin{equation}
\left(\sqrt{\frac{|\kappa_-|}{\kappa_+}}\right)_\text{quantum RN} =\frac{2n+1+p}{p}.
\end{equation}
The ratio of the two surface gravities being a rational number is a consequence of the quantization of the horizon area, as suggested in \cite{0102061,0209039}.

Since $m$ is an integer, the numerator must be divisible by $p$, so we write
\begin{equation}
2n+1 = lp,
\end{equation}
for some integer $l$. Furthermore, $2n+1$ is odd implies that \emph{both} $l$ and $p$ must be odd. We would then have
\begin{equation}\label{QMkp}
\frac{Q}{M}= \frac{2\sqrt{p}\sqrt{lp+p}}{lp+2p}=\frac{2\sqrt{l+1}}{l+2}, ~~l~\text{odd}.
\end{equation}

Thus the area quantization is consistent with $Q/M$ corresponding to the even-terms subsequence of the Kasner sequence, i.e.,
\begin{equation}\label{s1}
S_e:=\left\{\frac{Q}{M}\right\}_{l=1,3,\cdots}=\left\{\frac{2\sqrt{2}}{3}, \frac{4}{5}, \frac{2\sqrt{6}}{7}, \cdots\right\}
\end{equation}
The quantum dominance bound revealed by the maximum force argument corresponds to the first element of this subsequence.

But how do we know that the Kasner sequence is really relevant for black holes? Maybe this is just a coincidence? Though there remains this possibility, I would argue that it is not. There is another theorem of Kasner \cite{K1}, but in the context of ``irregular analytic curves'' (curves which, in a neighborhood of a given point $p$, taken as origin of orthogonal axes $(x,y)$, cannot be represented by setting $y$ equal to a series in integral powers of $x$, but can be represented by a series with fractional exponents):
\begin{theorem}\label{t1}
If the analytic curve is irregular at the point $P (0,0)$, with equation
\begin{equation}
y = \pm ix + C_{p+k-1}x^{\frac{p+k-1}{p}} + C_{p+k}x^{\frac{p+k}{p}}+ \cdots,
\end{equation}
then the limit of the arc-to-chord ratio is
\begin{equation}
L = \frac{2p}{2p+k+1}\sqrt{\frac{p+k-1}{p}}.
\end{equation} 
\end{theorem}
It is straightforward to check that this is \emph{exactly} of the form Eq.(\ref{Q-M}) for the quantized charge-to-mass ratio, if we were to identify $2n+1 = k-1$, or equivalently, $k=2(n+1)$. 
For $p=1$ this reduces to 
\begin{equation}
\frac{Q}{M}=\frac{2\sqrt{k}}{k+1}, ~~k~\text{even},
\end{equation}
which is the same result as before in Eq.(\ref{QMkp}), as we easily recognized that $k=l+1$. However, the point is that the quantized charge-to-mass ratio matches $L$ in the theorem even when\footnote{As Kasner wrote in \cite{K1}, ``the range of possible values of $L$ for the irregular arcs is greater, since the result depends on the two integers $p$ and $k$, characterizing the type of irregular (singular point) [sic]  and the contact with the minimal tangent. The set of numbers (in $L$) is dense everywhere between 1 and 0.'' } $p \neq 1$. This makes coincidence quite unlikely. 

The odd-terms subsequence of the Kasner sequence is the one complementary to the subsequence (\ref{s1})
\begin{equation}\label{o}
S_o:=\left\{\frac{Q}{M}\right\}_{l=2,4,\cdots}=\left\{1, \frac{\sqrt{3}}{2}, \frac{\sqrt{5}}{3}, \cdots\right\},
\end{equation}
the second term of which is the value of the Davies point (the first term is the extremal value). The elements of $S_o$ \emph{cannot} be equal to the values of quantized $Q/M$, but since $L$ in Theorem \ref{t1} is dense in the interval (0,1) there are always values of quantized $Q/M$ that is arbitrarily close to any given element of $S_o$. Since the Davies point corresponds to a divergence in the heat capacity, it is satisfying to see that \emph{an actual infinity cannot be attained by permissible values of quantized $Q/M$}, though of course the heat capacity can be arbitrarily large. We also see that an \emph{exactly extremal black hole is not admissible at the quantum level}. 

\section{Kerr Black Holes and Kasner Ratio}\label{kerr}

For the Kerr black hole, the situation is somewhat different. Its entropy can be quantized analogously as \cite{0211089}
\begin{equation}
S=\pi \left[2\left(n+j+\frac{1}{2}\right)\right], ~~n,j \in \Bbb{N}.
\end{equation}
However, since $S=\pi (r_+^2+a^2)$ instead of just $S=\pi r_+^2$, it seems that we cannot obtain the same result as before to match the Kasner sequence with the quantized $a/M$. Nevertheless, the value $a/M = 2\sqrt{2}/3$ does appear in a classical setting: the equatorial innermost stable circular orbit (ISCO) of a black hole that rotates at such a value of $a/M$ coincides with the ergosphere; any faster and the ISCO will drop below the ergosphere and so necessarily co-rotates with the black hole. Numerically the value $a/M = 2\sqrt{2}/3$ seems to also match the numerically obtained transition point for which superradiance (of massless scalar field) exceeds that of Hawking emission \cite{2306.17423}. In addition, $a/M=\sqrt{3}/2$ is also interesting in the Kerr case. Instead of the Davies point, it has to do with embedding condition: it is well-known that black holes with $a/M > \sqrt{3}/2$ cannot be isometrically embedded into $\Bbb{R}^3$ since the Gaussian curvature near the pole of the horizon surface (and nearby surfaces) becomes negative \cite{0706.0622,1708.07532}. Consequently, this leads to the divergence of the Wang-Yau quasi-local energy \cite{2406.10751} (see also \cite{1606.08177}). Curiously, as mentioned in the introduction, smooth embedding for the $(t,r)$-part of Reissner-Nordstr\"om geometry into (2+1)-dimensional Minkowski spacetime requires $Q/M < 2\sqrt{2}/3$ \cite{0305102}.

In \cite{MG}, the error ratio between the Hawking temperature and $T_\text{IRS}:=1/4\pi r_+$ associated with the inverse radius scale of the black hole was introduced. The quantity $T_\text{IRS}$ only agrees with Hawking temperature when $Q$ or $a$ vanishes. Otherwise the error ratio is
\begin{equation}
R(x):=\frac{T_\text{IRS}-T}{T}=\frac{1-\sqrt{1-x^2}}{\sqrt{1-x^2}},
\end{equation}
for the Kerr case ($x=a/M$), and similarly
\begin{equation}
R(x)=\frac{1}{2}\left(\frac{1-\sqrt{1-x^2}}{\sqrt{1-x^2}}\right),
\end{equation}
for the Reissner-Nordstr\"om case ($x=Q/M$).
One can also interpret $R(x)$ as the error ratio between the actual black hole temperature and the temperature of a putative Schwarzschild black hole whose areal radius is the same as the original black hole.
Essentially $R(x)$ is equivalent to a measure of the deviation between the classical horizon scale and the wavelength of the Hawking radiation \cite{MG}. 
For Reissner-Nordstr\"om black holes, $R(x) \geqslant 1$ if and only if $x \geqslant 2\sqrt{2}/3$; but for Kerr black holes, $R(x) \geqslant 1$ if and only if $x \geqslant \sqrt{3}/2$. Thus as far as the Hawking radiation is concerned, Kerr black holes do behave quite differently from the Reissner-Nordstr\"om case, therefore it is not a surprise that the Kasner ratio does not quite match the quantized $a/M$. 

The error ratio holds more surprises: for the Reissner-Nordstr\"om case, one has $R(\sqrt{3}/2)=1/2$, and indeed
\begin{equation}
R^\text{RN}(x)=\frac{1}{k} \Longleftrightarrow x = \frac{2 \sqrt{k+1}}{k+2}, ~k \in \Bbb{N},
\end{equation}
which we recognize as the Kasner ratio. The Kerr case does not work because 
\begin{equation}
R^\text{Kerr}(x)=\frac{1}{k} \Longleftrightarrow x = \frac{\sqrt{2k+1}}{k+1}, ~k \in \Bbb{N},
\end{equation}
which is not the same sequence. The fact that the error ratio of the Reissner-Nordstr\"om case is exactly $1/2$ of the Kerr case makes all the difference.
However, note that $R^\text{Kerr}$ is \emph{exactly} the odd subsequence of the Kasner sequence in Eq.(\ref{o}), i.e. the sequence complementary to the one that corresponds to the quantized horizon of Reissner-Nordstr\"om black hole (excluding the trivial case $k=0$, which is the first term of the Kasner ratio)! Why does a factor of $2$ in the error ratio of the Kerr case pick out the odd terms in the Kasner sequence? This remains obscure to me.

The Kasner sequence also shows up in the Kerr case in the classical context. First, let us express the Hawking temperature of a Kerr black hole as
\begin{equation}
T = \frac{1}{2\pi}(\kappa_\text{Sch} - {K}),
\end{equation}
in which $\kappa_\text{Sch}$ is the surface gravity of a Schwarzschild black hole of the same mass and ${K}$ essentially measures the deviation from the Schwarzschild case; c.f. the inverse radius scale ``temperature'' mentioned above which also measures deviation from being Schwarzschild.
($K$ is called the ``spring constant'' for Kerr black holes in \cite{1412.5432} and denoted as $k$ therein; see also \cite{2108.13435}). 
The dimensionless quantity $k_*:={K}/\kappa_\text{Sch}$ satisfies \cite{1412.5432}
\begin{equation}
\frac{a}{M} = \frac{2\sqrt{k_*}}{1+k_*},
\end{equation}
which is again the Kasner ratio; but because $k_* \in [0,1]$ \cite{1412.5432} it does not form a Kasner sequence. 
However, the Kasner ratio is symmetric under inversion, i.e., it is invariant under $k_* \mapsto 1/k_*$:
\begin{equation}
\frac{2\sqrt{\frac{1}{k_*}}}{1+\frac{1}{k_*}}=\frac{2k_*\sqrt{\frac{1}{k_*}}}{k_*+1}=\frac{2\sqrt{k_*}}{1+k_*},
\end{equation}
so the Kasner sequence can be formed after an inversion and restricting $k_*$ to $\Bbb{N}$.

Note that $k_* = a^2/r_+^2$ \cite{1412.5432}, and since $a^2=r_+r_-$, we have also the relation $k_*=r_-/r_+$. As we will soon see in the next section, indeed to obtain the Kasner ratio we do not need to involve the spring constant, since for Reissner-Nordstr\"om black holes we also have the same relation.

\section{Ratio of the Horizon Radii and The Inversion Symmetry of Kasner Ratio}\label{cc}
In the Reissner-Nordstr\"om case,  let us begin by re-writing the metric function $f(r)=1-\frac{2M}{r}+\frac{Q^2}{r^2}$ as \cite{0301173}
\begin{equation}
f(r) = \frac{(r-1)(r-\tilde{k})}{r^2},
\end{equation}
where $r=\tilde{k}$ marks the inner horizon and $r=1$ is fixed to correspond to the outer horizon. 
Then, because $M=(r_+ + r_-)/2$ and $Q^2=r_+r_-$, we have
\begin{equation}
M=\frac{1+\tilde{k}}{2},~~ Q=\sqrt{\tilde{k}},
\end{equation}
and thus the charge-to-mass ratio again yields the Kasner ratio
\begin{equation}
\frac{Q}{M} = \frac{2\sqrt{\tilde{k}}}{1+\tilde{k}}, ~~ \tilde{k}\in [0,1].
\end{equation}
Of course, this being a classical result, $\tilde{k}$ can take any real value between the unit interval (similar to $k_*$ in the preceding section), hence it does not distinguish special values like $2\sqrt{2}/3$ or $\sqrt{3}/2$.

Notice that (coincidentally?) the Kasner ratio is also the ratio between the geometric mean (GM) and the arithmetic mean (AM) of $\tilde{k}$ and 1, so the weak cosmic censorship bound $Q/M \leqslant 1$ is equivalent to the AM-GM inequality. In fact if we do not normalize $r_+=1$, we see that the mass and the charge are, respectively, the arithmetic mean and the geometric mean of $r_+$ and $r_-$ respectively\footnote{This can be checked directly, or by noting that the metric function $f(r)=0$ is equivalent to $r^2-2Mr+Q^2=(r-r_+)(r-r_-)=0$, and then the result follows from Vieta's formula.}. 
One may be tempted to say that since 
\begin{equation}\label{QM}
 \frac{Q}{M} = \frac{\sqrt{r_+r_-}}{(r_+ + r_-)/2}=\frac{\text{GM}}{\text{AM}},
\end{equation}
violating cosmic censorship (at least for Reissner-Nordstr\"om and Kerr black holes) amounts to violating the AM-GM inequality, which being a mathematical theorem, is impossible. This is however, incorrect for a few reasons. The first reason is this observation only applies to the exact Reissner-Nordstr\"om and Kerr solutions, and not their perturbations, so it does not amount to a proof for the weak cosmic censorship.
More importantly, the relations above for $Q$ and $M$ in terms of $r_+$ and $r_-$ still work if $r_+, r_- \in \Bbb{C}$, but the AM-GM inequality no longer applies. Therefore the AM-GM inequality is only a restatement of the fact that black hole solutions are restricted to the case $r_+,r_- \in \Bbb{R}$ (notice that $Q,M$ are still real numbers); it cannot explain why there is a cosmic censor.

Nevertheless, by writing $Q/M$ in the form of Eq.(\ref{QM}) we can appreciate why the Kasner ratio $2\sqrt{k}/(1+k)$ is invariant under inversion, i.e., under $k \longmapsto 1/k$. Eq.(\ref{QM}) is symmetric in $r_+$ and $r_-$, so if we divide both the numerator and denominator throughout by $r_+$, then
\begin{equation}\label{QM}
 \frac{Q}{M} = \frac{2\sqrt{\frac{r_-}{r_+}}}{1+\frac{r_-}{r_+}}.
\end{equation}
In this case $r_-/r_+$ takes the role of $k$, and it is bounded between 0 and 1. On the other hand, if we were to divide both the numerator and denominator throughout by $r_-$, then $r_+/r_-$ takes the role of $k$, and it can take any value equal to and above unity. 

Therefore we see that the Kasner ratio already appears in the classical setting, with $k$ playing the role of $r_+/r_-$ or its inverse. Consequently, the Kasner \emph{sequence}, with $k\in \Bbb{N}$, which appears in the quantum context, is the statement that $r_+$ is an integer multiple of $r_-$.

\section{Possible Relations to Graph Theory}\label{graph}

Are there any other evidence, or at least suggestive hints, that the Kasner ratio is relevant for understanding quantum aspects of black holes? For this we shall turn to graph theory.

There are various applications of random matrices to black holes, e.g., the late time behavior of horizon fluctuations in large anti-de Sitter black holes seems to be governed by random matrix dynamics \cite{1611.04650}. Random matrices also seem to govern the linear growth of complexity \cite{2106.02046}; it can also be applied to understand fuzzballs \cite{2301.11780}. Notably, random matrix models can be used to understand the spectral gap in the microstate spectrum of black holes \cite{2407.17583}, especially towards the extremal limit. Now, random matrices can be related to random graphs \cite{1910.08253}. 

Irrespective of random matrices and quantum information, random graphs are themselves interesting in the black hole contexts because, as noted in \cite{1805.08215}, generic random graphs have the property that a significant fraction of vertices in any connected region $A \subset V$, where $V$ is the vertex set, are located at the \emph{boundary} of $A$, much like a black hole whose number of degrees of freedom lies on the horizon area. In other words, random graphs may serve as toy models in the study of black hole microstates (see also \cite{1907.03090}). In \cite{1805.08215}, it was also shown that logarithmic scrambling time can be achieved in most quantum systems with sparse connectivity. Essentially, a sparse graph is a graph in which the number of edges is much less than the possible number of edges. This may shed some light on the scrambling of quantum information by a black hole. Indeed, sparseness would mean that not only black holes are fast scramblers \cite{0808.2096}, they are also efficient. 

Often, in addition to sparseness, we are also interested in the connectivity (i.e., not just the number of edges, but how well the vertices are connected). \emph{Expanders} are graphs that are very sparse but well-connected in some sense. For an application of expander graphs in an attempt to understand black holes as fast scramblers, see \cite{1204.6435}, in which it was argued that expander graphs provide a simple microscopic model for thermalization on quantum horizons. See also \cite{1306.3873}. For applications in the context of complexity, see \cite{1810.11563,2006.01280}.

Given a graph $G$ of $n$ vertices, the \emph{Cheeger constant} that quantifies the expansion of a graph, is defined as
\begin{equation}
h(G) := \min_{S\subset V, 0 <|s| \leqslant n/2} \frac{|\partial S|}{|S|},
\end{equation}
where $\partial S$ denotes the boundary of $S$ (the set of edges with exactly one endpoint in $S$). The Cheeger's inequality tells us that if $G$ is a $d$-regular graph and its adjacency matrix has eigenvalues $\lambda_1 \geqslant \lambda_2 \geqslant \cdots \lambda_n$, then the Cheeger constant is bounded:
\begin{equation}
\frac{d-\lambda_2}{2} \leqslant h(G) \leqslant \sqrt{2d(d-\lambda_2)}.
\end{equation}
The difference $d-\lambda_2$ is also called a spectral gap in this context. Thus it is interesting to know more about the bound on $\lambda_2$ for $d$-regular graphs. In fact, it is known that if the graph is random, then $\lambda_2 = 2\sqrt{d-1} + \varepsilon$ (where $\varepsilon>0$)  \cite{friedman1,friedman2}, and furthermore that if the graph is sufficiently large, then $\lambda_2 \geqslant 2\sqrt{d-1} - o_N(1)$ (where $o_N(1)$ refers to the asymptotic behavior as $n$ grows while $d$ is fixed). The latter result is known as the Alon–Boppana bound \cite{alon, nilli}. 

Can one achieve $\lambda_2 \leqslant 2\sqrt{d-1}$? The answer is yes, and these graphs are known as \emph{Ramanujan graphs}. (See also \cite{2203.07317} for dense random regular graphs.)
Note that $\lambda_1=d$ for $d$-regular graphs, so $\lambda_2$ can be understood as the largest eigenvalues less than $d$. For simplicity, we can just call it $\lambda$. That is, define
$\lambda:=\max_{|\lambda_i|<d} |\lambda_i|$, so that the Ramanujan graph is defined as those graphs that satisfy $\lambda \leqslant 2\sqrt{d-1}$. Note that if we normalize the eigenvalue so that $|\lambda_i|\leqslant 1$, then this amounts to dividing the eigenvalue by $\lambda_1$, so the definition for Ramanujan graph becomes
\begin{equation}
\lambda \leqslant \frac{2\sqrt{d-1}}{d}.
\end{equation}
This is the convention used in \cite{2110.01407,2302.07772}. With a slight shift of variable $d=k+1$, we recognize this as the Kasner ratio. Even if we do not work with the normalized eigenvalue, the (inverse) Kasner ratio still appears in another context: if $G$ is a $d$-regular Ramanujan graph, then its chromatic number\footnote{The smallest number of colors to color the vertices, such that adjacent vertices are colored differently.} is at least (see Corollary 1.5.4 of \cite{DSV})
\begin{equation}
\chi(G) \geqslant \frac{d}{2\sqrt{d-1}}.
\end{equation}

The case $d=3$ for which the normalized eigenvalue is bounded by $2\sqrt{2}/3$ (or $\chi(G)\geqslant 3/2\sqrt{2}$ without the normalization) corresponds to a $3-$regular graph (i.e., each vertex has 3 edges). Such graphs were utilized in
``quantum graphity'' models \cite{0801.0861,1409.2557}, a background independent approach for emergent geometry, in which space
is represented as a dynamical graph. Perhaps the occurrences of Kasner ratio in black hole physics can also be related to graph theoretical models of horizon dynamics. However, even so, it is unclear why the Kasner ratio appears as $Q/M$ and $a/M$. Does this mean that for certain values of $Q/M$ and $a/M$, the graph model for the black holes is Ramanujan? To answer this question, however, one has to first come up with graph models for charged and rotating black holes, which is in itself a difficult but interesting problem.

The number $2\sqrt{2}/3$ also appears when one considers the dynamics of random graph formation \cite{1110.1134}, with the leading singular component of a certain generating function taking the form $B(1-x)^{3/2}$. The prefactor $B$ tends to $2\sqrt{2}/3$ as the number of degrees $d \to \infty$.

\section{Discussion: More Questions and Musings}\label{discuss}

In this work, I pointed out that some special values of the charge-to-mass ratio $Q/M$ of Reissner-Nordstr\"om black holes, as well as the rotation parameter to mass ratio $a/M$ for Kerr black holes ($1, 2\sqrt{2}/3, \sqrt{3}/2$) are consecutive elements of the Kasner sequence, which originated from the arc-to-chord ratio of complex curves. Interestingly, Kasner has conjectured that the sequence may have to do with quantum line spectrum. However, as discussed in Sec.(\ref{3}), we should perhaps revise his conjecture into: 
\begin{conjecture}
The even elements of the Kasner sequence $\left\{\frac{2\sqrt{k}}{1+k}\right\}$ corresponds to special values of $Q/M$ in the quantized Reissner-Nordstr\"om black hole spacetimes when $r_+ = kr_-$, where $k \in \Bbb{N}$.
\end{conjecture}
In addition, we have seen that the odd sequence is not admissible. In particular this implies that that Davies point cannot be exactly attained, thus avoiding a truly divergent heat capacity. Similarly, an exactly extremal case is not admissible at the quantum level. This is also satisfying in view of known puzzles concerning extremal black holes (e.g., they have zero temperature but nonzero entropy), and is supported by the fact that under Hawking evaporation and Schwinger emission, black holes can approach but cannot attain extremality \cite{HW,1907.07490}.

Despite the calculations done in Sec.(\ref{3}), this is still just a conjecture because (1) we still do not have sufficient understanding of quantum gravity, so perhaps the quantization of the horizon area, and that of charge-to-mass ratio, still need to be revised; and more importantly (2) we need to derive/support this correspondence from a concrete quantum gravity theory or model, otherwise this may still just be a coincidence.

As mentioned, the quantized charge-to-mass ratio picks up the even terms in the Kasner sequence (the first of which is $2\sqrt{2}/3$), whereas the odd terms can be approached arbitrarily close if we consider irregular complex curves (the first nontrivial term being $\sqrt{3}/2$).
It should be remarked that there are also other differences between the odd and even terms. Namely, Kasner's ``Fundamental Theorem III'' \cite{K3} for complex \emph{surfaces}. For surfaces one can ask a similar question: given a point $P$ on the surface, and another point $Q$ moving along a regular path on the surface towards $P$, what value of the arc-to-chord ratio can be obtained? Call the set of values ``Kasner set''. Then,
\begin{theorem}
There are only 5 types of Kasner set for regular analytic surfaces:
\begin{itemize}
\item[(A)] The entire complex plane $\Bbb{C}$, including $0$ and $\infty$,
\item[(B)] All complex numbers with a single real exception; the exception being one of the odd terms of the Kasner sequence $R_3, R_5, \cdots$,
\item[(C)] The Kasner sequence,
\item[(D)] Just two real numbers: 1 and one of the even terms of the Kasner sequence $R_2, R_4, \cdots$,
\item[(E)] Just the real number 1.
\end{itemize}
\end{theorem}
Kasner further showed that the most general type given a random surface is type B, and among the type B, $R_3=\sqrt{3}/2$ is the most often omitted value. More importantly, we see from the theorem that there are distinctions between the odd terms and even terms once we consider complex surfaces. But if we were to match the Kasner sequence with quantized $Q/M$, then the distinctions appear already even for complex curves.

Perhaps the biggest mystery is: what do black holes have to do with complex curves? More specifically, why does the limit of arc-to-chord ratio of complex curve have to do with physical ratios of black holes ($Q/M$ and $a/M$)? Perhaps this may point us towards a better understanding of quantum black holes? Indeed, graph theory has provided some insights into the scrambling of quantum information and the area law of the horizon, and in graph theory one also finds a bound involving the Kasner ratio (the Ramanujan graph), so perhaps this hope is not too far-fetched and we can speculate that:
\begin{conjecture}
For a suitably constructed graph theoretic model of Reissner-Nordstr\"om and Kerr black holes, the graphs are Ramanujan when $Q/M$ and, respectively $a/M$, is a Kasner ratio.
\end{conjecture}
The main problem is that it is not clear what ``suitably constructed'' means, but perhaps this can serve as a guide for future model-building.

On the other hand, the fact that $2\sqrt{2}/3$ and $\sqrt{3}/2$ occur so many times in black hole physics, but hardly any other Kasner ratios (like ${4}/{5}, {\sqrt{5}}/{3},$ and ${2\sqrt{6}}/{7}$), also requires some explanations. For the Reissner-Nordstr\"om case, being related to area quantization, perhaps only the first few terms in the Kasner sequence affect black hole thermodynamics in an obvious way, just like only the first few hydrogen emission spectrum lines are visible. But how then can we explain the absence of these values in the Kerr case? This might point towards either a different physical origin for the ratios $2\sqrt{2}/3$ and $\sqrt{3}/2$ in the Kerr case, or that there exist some deeper mathematical relations between these two ratios (regardless of Kerr or Reissner-Nordstr\"om). The latter possibility will be pursued in a forthcoming study.

It is worth mentioning that bounds involving $2\sqrt{2}/3$ also occur in various contexts that have nothing to do with black holes, even in pure mathematics. Consider the unit disk $\Bbb{D}$ in the complex plane, and define the pseudohyperbolic distance between $z,w\in \Bbb{D}$ by
\begin{equation}
\rho(z,w):=\left|\frac{z-w}{1-\bar{w}z}\right|,
\end{equation}
where a bar denotes complex conjugation.
Let $S=\left\{z_i\right\}$ be a sequence in $\Bbb{D}$ with 
\begin{equation}
\delta_S := \inf_j \prod_{k\neq j} \rho(z_j, z_k) > 0.
\end{equation}
The quantity $\delta_S$ is called the \emph{separation constant}.
If $\left\{w_i\right\}$ is a bounded sequence in $\Bbb{C}$, then there exists a bounded analytic function $F \in H^\infty$ (the Hardy space $H^\infty$ is the vector space of bounded holomorphic functions on the unit disk) with $\lVert F \rVert_{H^\infty} \leqslant \mathcal{M}(\delta_S) \sup_j |w_j|$, where
\begin{equation}\label{Md}
\mathcal{M}(\delta) := \frac{1}{\delta^2}\left(1+\sqrt{1-\delta^2}\right)^2,
\end{equation}
that solves the interpolation problem $F(z_j)=w_j$ for all $j$ \cite{2006.04901}. We say that $\left\{z_i\right\}$ is an interpolating sequence.
Let $\lVert \cdot \rVert$ denotes the 2-norm for vectors and the corresponding operator norm for matrices: $\lVert A \rVert:=\sup_{\lVert A \rVert=1}\lVert Ax \rVert$.
Consider an $n \times n$ matrix $A$ with distinct eigenvalues and satisfying $\lVert A \rVert \leqslant 1$. Suppose that its spectrum $\sigma(A) \subset \Bbb{D}$, then for any $f\in H^\infty$ , we have the bound
\begin{equation}
\lVert f(A) \rVert\leqslant \mathcal{M}(\delta_A) \max_{z\in \sigma(A)} |f(z)|,
\end{equation}
where $\delta_A \equiv \delta_{\sigma(A)}$.
For $\delta_A \geqslant 2\sqrt{2}/3$, we have $\mathcal{M}(\delta_A) \leqslant 2$, so that matrices with well-separated eigenvalues satisfy the \emph{Crouzeix's conjecture} \cite{2006.04901} (see also \cite{2112.06321}) first proposed in \cite{crouzeix}, which states that, for any polynomial $f$ and any square matrix $A$, the operator norm of $f(A)$ satisfies
\begin{equation}
\lVert f(A) \rVert \leqslant 2 \sup\left\{|f(z)|: z \in W(A)\right\},
\end{equation}
where $W(A)=\left\{\left\langle Ax,x\right\rangle: \lVert x \rVert\leqslant 1\right\}$ is the numerical range of $A$.

Note that the term inside the bracket of $\mathcal{M}(\delta)$ in Eq.(\ref{Md}), namely $1+\sqrt{1-\delta^2}$, is of the form of the event horizon radius for Kerr and Reissner-Nordstr\"om black holes, so mathematically speaking the fact that $2/\sqrt{2}/3$ appears is perhaps not a coincidence. In fact, for the Reissner-Nordstr\"om case, the somewhat mysterious expression of $\mathcal{M}$ has a \emph{physical} meaning, it is the ratio $r_+^2/Q^2$ or $S/S_\text{ext}$, i.e., the ratio of the black hole entropy to the extremal black hole entropy. Furthermore, if we consider $\delta$ as the charge-to-mass ratio and substitute in the quantized version given by Eq.(\ref{Q-M}), then the function $\mathcal{M}$ becomes
\begin{equation}
\mathcal{M}(Q/M)|_\text{quantum RN} = \frac{2n+p+1}{p}, 
\end{equation}
which is exactly the integer $m$ in Eq.(\ref{mnumber}) that generates the even Kasner subsequence! In fact, ${(2n+p+1)}/{p}=S/S_\text{ext}$ agrees with the result in \cite{1111.4164} that ${(2n+p+1)}/{p}$ equals the quantized version of $\sqrt{A_+/A_-}$, where $A_-$ denotes the area of the inner horizon (since $Q^2=r_+r_-$). This suggests that we should be able to find the other Kasner ratios (not just $2\sqrt{2}/3$) to have some significance in the context of complex interpolation problem. This is indeed the case.

The expression $\mathcal{M}(\delta)$ already appeared in a paper by Earl \cite{earl} from 1970, in which we also see how the Kasner ratios arise in complex interpolation problem.
Let $\left\{z_i\right\}$ be an interpolating sequence, Earl proved that if the sequence has at least 3 elements, then the disk defined by
\begin{equation}
\rho(z,z_n) \leqslant \tau,
\end{equation}
 are disjoint for $\tau \leqslant {\mathcal{M}(\delta)^{-\frac{1}{2}}}$. On the other hand, 
\begin{equation}
\rho(z_m,z_n) \leqslant \frac{2\tau}{1+\tau^2}.
\end{equation}
The expression on the RHS has the form of Kasner ratio upon the obvious substitution\footnote{This bound comes about because $\mathcal{M}= \frac{1}{\delta^2}\left(1+\sqrt{1-\delta^2}\right)^2 \Longrightarrow \delta = 2\sqrt{\mathcal{M}^{-1/2}}/(1+\mathcal{M}^{-1/2}).$ } $\tau \mapsto \sqrt{k}$.

It might be interesting to investigate further to see if complex interpolation has deeper physical interpretations related to Reissner-Nordstr\"om black holes. That is to say, although these two are unrelated problems, maybe the underlying mathematical connections can help us to understand both problems from different perspectives. 

Another instance where some Kasner ratios occur is in the context of univalent polynomials, i.e., holomorphic polynomials that are injective. Let $\mathcal{P}_n$ denotes the class of normalized polynomials of the form $P_n(z) = z + a_2 z^2 + \cdots + a_n z^n$,
which are univalent in $|z| < 1$. The sufficient and necessary condition for $P_2 \in \mathcal{P}_2$ is that $|a_2| \leqslant 1/2$. For $P_3 \in \mathcal{P}_3$, suppose $a_2, a_3 \in \Bbb{R}$, then it can be shown that \cite{brannan} $a_2 \leqslant 1/2, 4/5, 2\sqrt{2}/3$, if $a_3 \leqslant 0, 1/5, 1/3$, respectively. The numbers $4/5$, and $2\sqrt{2}/3$ are both Kasner ratio.

Recall that in Sec.(\ref{2}) we mentioned that Kasner realized long ago that the Kasner ratio $R:=2\sqrt{n}/(n+1)$ implies \cite{K3}
\begin{equation}
\frac{R^2}{4}=\frac{m-1}{m^2}=\frac{1}{m}-\frac{1}{m^2}, 
\end{equation}
where $m=n+1$. 
Note that if we extend $n$ to complex number by setting $n=-z$, then this gives
\begin{equation}
-\frac{R^2}{4}=\frac{z}{(1-z)^2}.
\end{equation}
The factor $-1/4$ on the LHS is there for a reason, as 
on the RHS we recognize the \emph{Koebe function} whose image domain is $\mathbb{C} \setminus (-\infty, -{1}/{4}]$. This is the function that demonstrates that the factor $1/4$ in the ``Koebe $1/4$ theorem'' cannot be improved. The theorem, conjectured in 1907 by Paul Koebe, and proved in 1916 by Ludwig Bieberbach, states that the image of an injective analytic function $f: \Bbb{D} \to \Bbb{C}$ contains the disk whose center is $f(0)$ and whose radius is $|f'(0)|/4$. (The theorem is also relevant to the study of the aforementioned univalent polynomials; see \cite{1812.08311}). 

To conclude, in \cite{K3}, Kasner made the observation that Eq.(\ref{ls}) 
has a superficial resemblance to the Balmer series or Rydberg formula in general. He then boldly conjectured that the ratio may one day be applied to some quantum ``spectrum line series''. This is all the more remarkable given that his results were purely in the field of complex analysis or geometry, and there was no hint whatsoever that it is related to anything quantum mechanical. In this work, however, we have seen that the Kasner ratio might indeed be relevant to quantum mechanics of \emph{black holes}, including perhaps thermalization and scrambling of quantum information (via the possible relations to graph theoretical models). 
The possible connections to other complex analysis problems are also interesting.
Despite the circumstantial evidence, these are nevertheless just a revised version of Kasner's original conjecture; more studies are required to check if any of these make sense; or just a bunch of coincidences.

\end{document}